\documentclass[twocolumn,amsmath,amssymb,prb]{revtex4}

\usepackage{graphicx} 
\usepackage{dcolumn}  
\usepackage{bm}       

\begin{document}

\title{\textit{Ab initio} wavefunction based methods for excited states in solids:
correlation corrections to the band structure of ionic oxides}

\author{L. Hozoi, U. Birkenheuer, and P. Fulde}
\affiliation{Max-Planck-Institut f\"{u}r Physik komplexer Systeme,
             N\"{o}thnitzer Str. 38, 01187 Dresden, Germany}

\author{A. Mitrushchenkov and H. Stoll}
\affiliation{Universit\"{a}t Stuttgart, Pfaffenwaldring 57, 70550 Stuttgart,
             Germany}

\date{\today}

\begin{abstract}
\textit{Ab initio} wavefunction based methods are applied to the study
of electron correlation effects on the band structure of 
oxide systems.
We choose MgO as a prototype closed-shell ionic oxide.
Our analysis is based on a local Hamiltonian approach and performed on
finite fragments cut from the infinite solid.
Localized Wannier functions and embedding potentials are obtained from
prior periodic Hartree-Fock (HF) calculations.
We investigate the role of various electron correlation effects in
reducing the HF band gap and modifying the band widths. 
On-site and nearest-neighbor charge relaxation as well as long-range
polarization effects are calculated.
Whereas correlation effects are essential for computing accurate band
gaps, we found that they produce smaller changes on the HF
band widths, at least for this material.
Surprisingly, a broadening effect is obtained for the O $2p$ valence bands.
The \textit{ab initio} data are in good agreement with the energy gap and
band width derived from thermoreflectance and x-ray photoemission experiments.
The results show that the wavefunction based approach applied here
allows for well controlled approximations and a transparent identification
of the microscopic processes which determine the electronic
band structure.  
\end{abstract}

\pacs{71.15.-m,71.15.Qe,71.20.-b}

\maketitle

\section{Introduction}

The proper treatment of electron correlation effects in molecules 
and solids stands at the heart of modern electronic structure
theory 
\cite{fulde_95,march_96,fulde_wf_02,mcweeny_96,QC_book_00}.
In the study of molecular systems at least,
wavefunction based quantum chemistry is known to provide a rigorous
theoretical framework for addressing the electron correlation problem.
The standard quantum chemical methods \cite{mcweeny_96,QC_book_00} make
possible the construction of approximate wavefunctions at 
levels of increasing sophistication and accuracy and offer thus a systematic 
route to converged results.
Advanced wavefunction based calculations can be routinely
performed nowadays for small and medium size molecules.
However, algorithms able to treat electron correlation effects
in periodic systems are still at their infancy. 
The simplest correlation method is based 
on the M\o ller-Plesset equations and  
second-order M\o ller-Plesset (MP2) perturbational schemes for solids 
were actually implemented by several groups
\cite{suhai_pt2_83,liegener_pt2_88,ladik_poly_99,bartlett_pt2_96,hirata_pt2_98,
reinhardt_pt2_98,ayala_pt2_01,pisani_pt2_05,ayjamal_pt2_cc_02}.
As an extension of MP2, \textit{ab initio} many-body Green's function
techniques were developed too
\cite{hirata_Gf_00,pino_Gf_04,albrecht_Gf_01,albrecht_02,buth_Gf_05}.
In addition, investigations based on coupled-cluster (CC) theory were
initiated
\cite{ayjamal_pt2_cc_02,foerner_cc_92,foerner_cc_97,fink_cc_95,hirata_cc_04}.

The main point when applying quantum chemical methods to correlation
calculations is to make use of the local character of the correlation
hole which is surrounding an electron.
The latter optimizes the Coulomb repulsion between 
electrons and its accurate description is the essence of the
correlation treatment.
Starting point is a Hartree-Fock (HF) calculation on top of 
which the correlation calculations are implemented.
For insulators and semiconductors, the development of efficient 
computational tools is facilitated by the use of optimally
localized Wannier functions.
Several orbital localization procedures were proposed in the context
of periodic HF calculations.
They normally rest on the \textit{a posteriori} transformation of
the crystal Bloch orbitals, see for example Refs.
\onlinecite{suhai_pt2_83,pisani_pt2_05,foerner_cc_92,foerner_cc_97,fink_cc_95,%
zico_01,birken_cb_05}. 
Alternative approaches were elaborated by Shukla \textit{et al.}
\cite{shukla_98,shukla_99,ayjamal_LiH_00,sony_06} and in the Toulouse
group \cite{reinhardt_pt2_98,malrieu_WOs_97}, where the self-consistent-field 
(SCF) equations are solved directly in the Wannier representation.
 
Local correlation methods such as the incremental scheme of Stoll
\cite{stoll_92,paulus_rev06} and the local Hamiltonian formalism of Horsch 
\textit{et al.} \cite{horsch_84} and Gr\"{a}fenstein \textit{et al.}
\cite{grafen_93,grafen_97,albrecht_00} were applied before to the 
rigorous determination of both ground-state
\cite{ayjamal_pt2_cc_02,shukla_98,shukla_99,ayjamal_LiH_00} 
and excited-state \cite{albrecht_02,bezugly_04,birken_06,CFLH_06,pahl_xx} 
properties of infinite systems.
Materials under investigation were crystalline LiH, LiF, and LiCl
\cite{shukla_98,shukla_99,ayjamal_LiH_00,albrecht_02}, diamond \cite{birken_06,pahl_xx}, 
silicon \cite{birken_06}, beryllium \cite{CFLH_06}, carbon, and boron-nitrogen chains 
\cite{ayjamal_pt2_cc_02}, and \textit{trans}-polyacetylene
\cite{bezugly_04}.
In the present paper we extend these studies to the case of ionic
oxide compounds and choose MgO as a prototype and, at the same time, 
relatively simple insulating oxide.
We compute correlation induced corrections to the valence and 
conduction HF energy bands and find that our final estimate for the
fundamental gap is in rather good agreement with the experimental
data.
Our analysis provides also a clear picture of the major correlation
effects that are responsible for the reduction of the HF band gap. 
The present study on magnesium oxide should open the way to similar 
calculations for more complex materials such as the $3d$ 
transition-metal oxides.

\section{Computational approach}

Periodic HF calculations were carried out for crystalline
MgO with the \textsc{crystal} package \cite{crystal}.
Wannier functions associated with the HF valence and low-lying 
conduction bands were determined with the Wannier-Boys localization
module \cite{zico_01} of the same program.
For conduction-band states, this may be a somewhat tedious task.
However, localized Wannier orbitals can also be derived
in more difficult cases with band disentanglement techniques
\cite{birken_cb_05}. 

For computing the correlation induced corrections to the valence and
conduction bands we adopt the same quasiparticle picture and local
Hamiltonian formalism as employed in Refs.
\onlinecite{horsch_84,grafen_93,grafen_97,albrecht_00,bezugly_04,birken_06,CFLH_06,pahl_xx}.
The physical reasoning is as follows.
When an electron (or hole) is added to the $N$-particle system,
its surroundings relax and polarize due to the additional charge.
This response of the system lowers the energy it takes to add the
extra electron (or hole).
The particle plus the modified surroundings move together
through the system in the form of a Bloch wave and define a
quasiparticle.
At the same time, some of the correlation contributions which are
present in the ground-state of the $N$-particle system are no
longer operative in the $(N\!+\!1)$ [or $(N\!-\!1)$] -particle system, because some
excitations are now blocked.
This effect is called loss of ground-state correlation and must be 
accounted for too \cite{horsch_84,birken_06}. 
A basic feature of our approach is that the correlation treatment is
performed on a finite cluster $\mathcal{C}$ cut from the extended
periodic system.
With the cluster-in-solid embedding technique elaborated 
by Birkenheuer \textit{et al.} \cite{birken_emb_1,birken_emb_2,birken_06},
the cluster $\mathcal{C}$ is divided into an active region
$\mathcal{C}_A$ that supports the local occupied and virtual orbitals
entering the post-HF calculations and a spatial ``buffer'' domain
$\mathcal{C}_B$ including a number of atomic sites
(and basis functions) whose role is to provide a good representation
for the tails of the Wannier-like orbitals of the active region
$\mathcal{C}_A$.
Thus $\mathcal{C} = \mathcal{C}_A + \mathcal{C}_B$.
All Wannier orbitals centered in $\mathcal{C}_B$
(and also the environment)
will be held frozen in the correlation calculations.

The orbital set associated with the finite-size cluster is obtained
from the data supplied by the Wannier-Boys localization module of
\textsc{crystal} \cite{zico_01}. 
The interface program \cite{crystal_molpro_int} written for this
purpose yields in addition an embedding potential corresponding to
the frozen environment, i.e., the surrounding HF electron sea. 
The detailed procedure for constructing the embedded cluster and
the associated orbital set is described elsewhere \cite{birken_emb_1,birken_emb_2}.
Here we merely give an outline of this.
The core, valence, and low-lying conduction-band cluster orbitals
are generated by projecting the crystal Wannier functions onto the
set of atomic basis functions attached to the cluster region
$\mathcal{C}$,
\begin{equation}
|w_n^\prime(\mathbf{R}) \rangle = \sum_{\beta,\alpha\in\mathcal{C}} |\beta\rangle
                                S_{\beta\alpha}^{-1} \langle\alpha|
                                w_n(\mathbf{R})\rangle \,, 
\end{equation} 
where $\alpha$ and $\beta$ are Gaussian basis functions centered 
within the region $\mathcal{C}$, $S_{\beta\alpha}^{-1}$ represents the
inverse overlap matrix for the basis set attached to $\mathcal{C}$,
$|w_n(\mathbf{R})\rangle$ is a Wannier orbital with index
$n$ and centered in the unit cell corresponding to the lattice
vector $\mathbf{R}$, and $|w_n^\prime(\mathbf{R})\rangle$ is its
projected counterpart expressed exclusively in terms of basis 
functions centered in $\mathcal{C}$.
The $|w_n^\prime(\mathbf{R})\rangle$ functions are neither normalized
nor are they orthogonal to each other because of the projection
procedure mentioned above.
Therefore they are group-wise orthonormalized in the following order:
active core, active occupied, active low-lying conduction-band
orbitals, buffer core, buffer occupied, buffer low-lying conduction-band
orbitals.
This way the contamination of the most important types of orbitals
is minimized.
Group-wise orthonormalization actually means to use Schmidt orthogonalization
for the inter-group orthogonalization and L\"{o}wdin orthonormalization
inside the groups. 
This set of orthonormal orbitals will be denoted by
$|\tilde{w}_n^\prime(\mathbf{R})\rangle$.
For the construction of the variational space to be used in the subsequent
correlation calculations we follow the prescription suggested by
Pulay and Saeb\o \ \cite{pulay_PAOs}. 
So-called virtual projected atomic orbitals (PAO's) are generated from the
Gaussian basis functions associated with the {\it active} region $\mathcal{C}_A$ 
by projecting out the occupied and the low-lying conduction-band orbitals
$|\tilde{w}_n^\prime(\mathbf{R})\rangle$
via a Schmidt orthogonalization scheme.
Thereafter, the PAO's are L\"{o}wdin orthonormalized among themselves
to facilitate their subsequent use in the correlation calculations.

By making the buffer region $\mathcal{C}_B$ sufficiently large
the original Wannier orbitals of $\mathcal{C}_A$ are well
represented.
For example, for the Wannier orbitals corresponding to the lowest
four conduction bands of MgO and centered at a Mg site,
the six ligands surrounding the given Mg ion should be included
in $\mathcal{C}_A$ because large contributions arise
not only from the Mg $3s$ and $3p$ basis functions but 
also from the 
nearest-neighbor (NN) oxygen $2s$,$3s$ and $2p$,$3p$
components.
Farther neighbors need not be included in the central region
$\mathcal{C}_A$ but may be put into the buffer zone $\mathcal{C}_B$ 
because the weight of the longer-range tails is small.
Whereas a high-quality description can be easily achieved for the
Wannier orbitals centered in the active region, the 
representation of the Wannier functions in the buffer zone is 
less accurate.
The impact of this deficiency on the correlation calculations, however, is
completely compensated by an appropriate choice of the embedding potential.
Actually, the Gaussian orbital representation $V^{\rm emb}_{\alpha\beta}$
of the embedding potential is constructed by
\begin{equation}\label{eqVemb}
V^{\rm emb}_{\alpha\beta} = F^{\rm crys}_{\alpha\beta}
                          - F[P^{\cal C}]_{\alpha\beta} \,, \qquad \alpha,\beta\in{\cal C}
\end{equation}
where $F^{\rm crys}$ is the self-consistent Fock operator from the periodic
HF calculation and $F[P^{\cal C}]$ is the Fock operator associated with the
density operator
\begin{equation}\label{eqPfroz}
P^{\cal C} = 2 \sum_\nu^{\rm occ} | \tilde{w}_\nu^\prime \rangle
                                  \langle \tilde{w}_\nu^\prime |
\end{equation}
arising from all occupied orbitals $|\tilde{w}_\nu^\prime\rangle$ which enter
the subsequent correlation calculations explicitly. This way, the correlation
calculations are effectively performed in an {\it infinite} frozen
HF environment.

The data concerning the occupied and virtual orbitals of the cluster  
is transferred via the \textsc{crystal-molpro} interface \cite{crystal_molpro_int}
to the quantum chemistry program \textsc{molpro} \cite{molpro_2006}. 
The same holds for the matrix representation $F^{\rm crys}_{\alpha\beta}$
of the self-consitent Fock operator of the periodic host system. The embedding
potential itself is constructed according to Eqs. (\ref{eqVemb}) and
(\ref{eqPfroz}) using the MATROP module of the \textsc{molpro}
program package.

Local electron removal and electron addition one-particle
configurations can be defined in terms of the set of 
occupied and virtual orbitals localized within the spatial domain
$\mathcal{C}_A$:
\begin{equation}\label{eqPhiNpm1}
|\Phi_{\mathbf{R}p\sigma}^{N-1}\rangle = c_{\mathbf{R}p\sigma}
                                             |\Phi\rangle \ \ \ 
\mathrm{and} \ \ \ 
|\Phi_{\mathbf{R}q\sigma}^{N+1}\rangle = c_{\mathbf{R}q\sigma}^{\dagger}
                                             |\Phi\rangle \,,
\end{equation}
where $c_{\mathbf{R}p\sigma}$ and $c_{\mathbf{R}q\sigma}^{\dagger}$
are annihilation and creation operators for the valence and conduction
band $\sigma$-spin orbitals $|\tilde{w}_p^\prime(\mathbf{R})\rangle$ and
$|\tilde{w}_q^\prime(\mathbf{R})\rangle$, respectively, and 
$|\Phi\rangle$ is the single-determinant ground-state wavefunction
of the $N$-electron system.
For clusters which are large enough, the Hartree-Fock valence and
conduction energy bands of the periodic crystal can be
recovered by diagonalizing $\mathbf{k}$-dependent matrices of 
the following form:
\begin{equation}\label{eqHkHF}
H_{nn^\prime}^{\mathrm{HF}} (\mathbf{k}) = \sum_{\mathbf{R}}
                                           e^{i\mathbf{k}\mathbf{R}}
\langle \Phi_{\mathbf{0}n       \sigma}^{N\mp 1} | H - E_0^{\mathrm{HF}} |
        \Phi_{\mathbf{R}n^\prime\sigma}^{N\mp 1}    \rangle \,.
\end{equation}
Here $\mathbf{0}$ stands for the reference unit cell and
$E_0^{\mathrm{HF}}$ denotes the ground-state HF energy of the neutral
$N$-electron system. 
The diagonal terms $\mathbf{R}\!=\!\mathbf{0}$, $n^\prime\!=\!n$
in this expression are directly related to the on-site excitation energies within the 
Koopmans approximation \cite{mcweeny_96,QC_book_00}, i.e.,
ionization potentials
\begin{equation}
\mathrm{IP}_{pp}^{\mathrm{HF}}(\mathbf{0}) = 
\langle \Phi_{\mathbf{0}p\sigma}^{N-1} | H |
        \Phi_{\mathbf{0}p\sigma}^{N-1} \rangle - E_0^{\mathrm{HF}} =
- \epsilon_{\mathbf{0}p}^{\mathrm{HF}} > 0 
\end{equation} 
and electron affinities
\begin{equation}
\mathrm{EA}_{qq}^{\mathrm{HF}}(\mathbf{0}) = E_0^{\mathrm{HF}} -
\langle \Phi_{\mathbf{0}q\sigma}^{N+1} | H |
        \Phi_{\mathbf{0}q\sigma}^{N+1} \rangle =
- \epsilon_{\mathbf{0}q}^{\mathrm{HF}} \,,
\end{equation}                    
with the latter being negative in the case of MgO.
The off-diagonal terms are the hopping matrix elements in a tight-binding
representation,
\begin{equation}\label{eqtnmHF}
t_{nn^\prime}^{\mathrm{HF}} (\mathbf{R}) = 
\langle \Phi_{\mathbf{0}n       \sigma}^{N\mp 1} | H-E_0^{\mathrm{HF}} |
        \Phi_{\mathbf{R}n^\prime\sigma}^{N\mp 1} \rangle \ .
\end{equation}

\begin{figure}[!b]
\includegraphics*[angle=270,width=1.0\columnwidth]{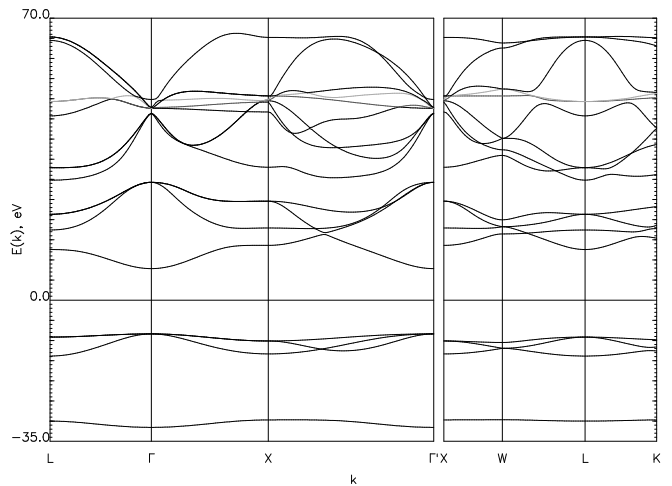}
\caption{Hartree-Fock band structure of bulk MgO.
The core O $1s$ and Mg $1s-2p$ bands are not shown in the figure.}
\end{figure}

Starting from the Hartree-Fock energy bands we include next the
effects of electron correlations.
Within the quasiparticle approximation we may introduce for the
($N\!+\!1)$-particle system a wavefunction of the following form:
\begin{equation}
|\Psi_{\mathbf{R}q\sigma}^{N+1}\rangle =
\exp{(S)}\, c_{\mathbf{R}q\sigma}^{\dagger} |\Psi\rangle \,,
\end{equation} 
where $|\Psi\rangle$ is the ground-state of the $N$ electron system.
The operator $S$ does not commute with $c_{\mathbf{R}q\sigma}^{\dagger}$
and describes the correlation hole.
For a more detailed discussion we refer to Ref. \onlinecite{fulde_95}.
Alternatively, we may write 
$|\Psi_{\mathbf{R}q\sigma}^{N+1}\rangle =
\Omega\, c_{\mathbf{R}q\sigma}^{\dagger} |\Phi\rangle$, 
where $|\Phi\rangle$ is the SCF ground-state and $\Omega$ acts like
a wave- or Moeller operator \cite{fulde_wf_02}.
Therefore, when correlations are taken into account, the analogue of 
(\ref{eqHkHF}) becomes
\begin{equation}\label{eqHkcorr}
H_{nn^\prime}(\mathbf{k}) =  \sum_{\mathbf{R}}
                            e^{i\mathbf{k}\mathbf{R}}
\langle \Psi_{\mathbf{0}n       \sigma}^{N\mp 1} | H - E_0 |
        \Psi_{\mathbf{R}n^\prime\sigma}^{N\mp 1}  \rangle \,,
\end{equation}
where $E_0$ is the energy of the $N$-particle correlated
ground-state $|\Psi\rangle$ (see also Refs.
\onlinecite{grafen_93,grafen_97,albrecht_00,bezugly_04,birken_06}).
The correlation hole around an added electron (or hole) consists of a 
short-range and a long-range part.
The short-range part originates from 
intra-atomic and short-range interatomic relaxation and 
polarization effects.
We construct the short-range part of the correlation hole by separate
orbital optimizations.
Thereby the Wannier orbital to which the extra electron (hole) is
attached is kept frozen and the changes within a finite region around
it are determined by performing an additional SCF calculation \cite{birken_06,CFLH_06}.
This is sufficient if electron correlations are weak or moderate.
If they are strong we would have to take into account the
fact that changes in the nearby surroundings are
decreased when correlation effects among the electrons in that neighborhood
are accounted for. 
The long-range part of the correlation hole consists of long
range polarization of the environment.
The effect of long-range polarization on the diagonal
Hamiltonian matrix elements is estimated in this paper by applying
the approximation of a dielectric continuum.
With the known dielectric constant of MgO, $\epsilon_0\!=\!9.7$,
we can directly determine the polarization energy of a charge $\pm e$
outside a sphere of radius $R$.
In addition, differential correlation effects related to the existence of a
different number of electrons in the system's ground-state and
in the $(N\!\pm\!1)$ excited states are investigated by
subsequent configuration-interaction (CI) calculations.

\section{Correlation corrections to the band structure of MgO}

Magnesium oxide is a prototype closed-shell ionic material that 
crystallizes under normal conditions in the rocksalt structure.
It is extensively used in materials science as substrate 
for the epitaxial growth of films of other compounds.
In recent years, it has attracted renewed interest because of 
its use as tunnel barrier in magnetic tunnel junctions
\cite{MTJs_MgO_04}.
A different area where MgO attracted attention is optoelectronics.
It turns out that when alloyed with ZnO, depending on the precise
chemical composition, the band gap of the system can be tuned 
along an interval ranging from 3.3 to 7.8 eV  
\cite{MgZnO_choopun_02}. 
The latter number represents the fundamental gap of MgO
\cite{gap_MgO}.

\begin{figure}[!t]
\includegraphics*[angle=90,width=0.80\columnwidth]{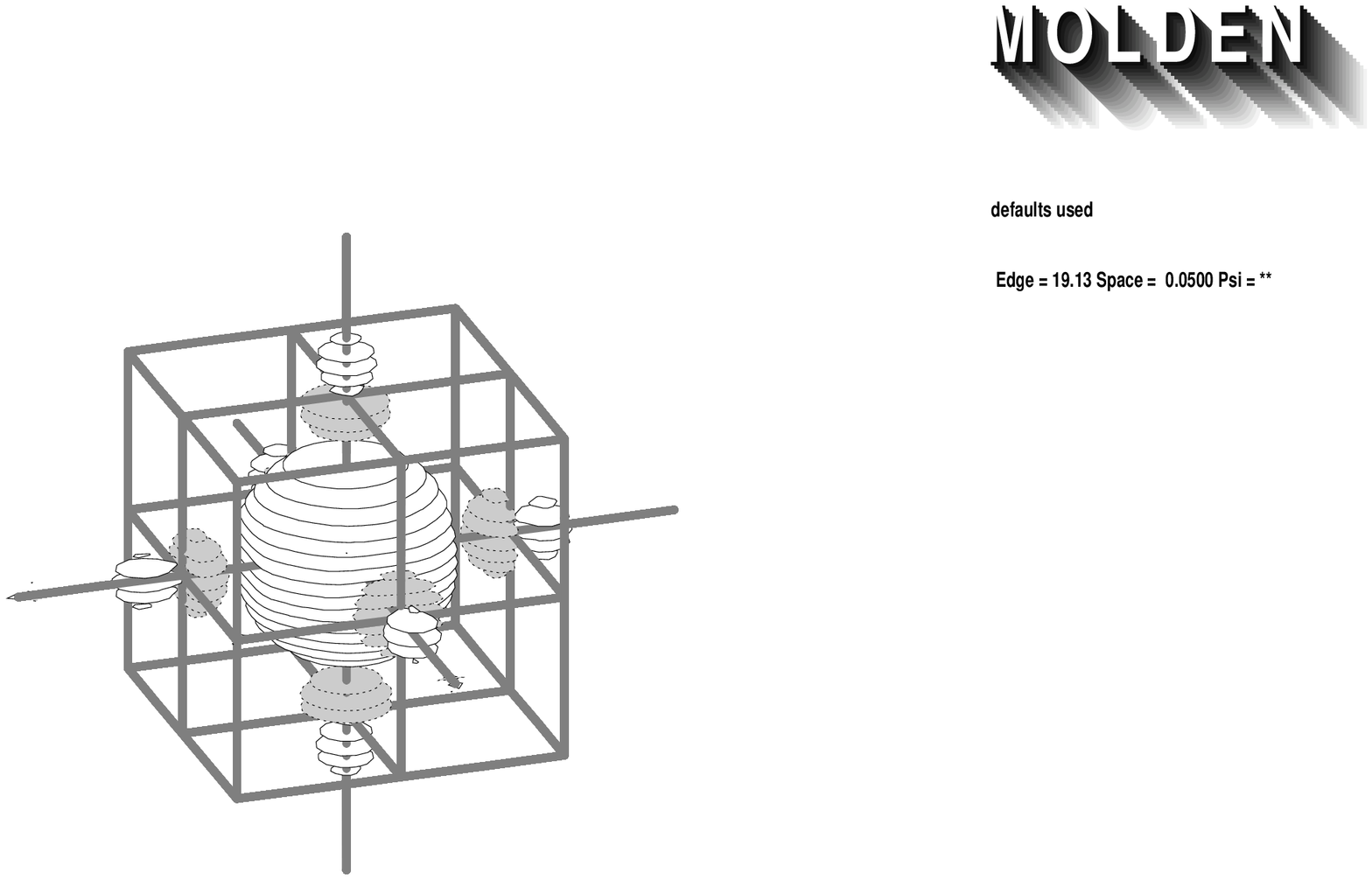}
\caption{Plot of the Mg $3s$-like conduction-band Wannier orbital
after projection onto a [Mg$_{19}$O$_{14}$] cluster.
There is substantial weight at the NN oxygen sites.}
\end{figure}

The valence bands of MgO have oxygen $2p$ character, whereas
the lowest conduction bands are mainly related to the Mg $3s$
and $3p$ orbitals, with some admixture from the O $2s$,$3s$
and $2p$,$3p$ functions.
The Hartree-Fock energy bands are shown in Fig.1.
We employed for our calculations the lattice constant reported by
Sasaki \textit{et al.}, $a\!=\!4.217$\,\AA \ \cite{mgo_LatCst_79}, and 
Gaussian-type basis sets from the standard \textsc{crystal} library. 
Basis sets of triple-zeta quality, Mg 8-511 \cite{BSs_MgO_92},
were used for the cations and triple-zeta basis sets 
suplemented with polarization functions, O 8-411$^{\ast}$
\cite{BSs_MgO_92}, were applied for the more polarizable O
ions.
At the Hartree-Fock level and with this choice of the basis
functions, the fundamental gap of the system is 16.20 eV,
i.e., 8.4 eV larger than the experimental value.
As illustrated in Fig.1, there is a separation of about 16
eV between the O $2s$ and O $2p$ bands.
There is also a clear separation between the low-lying Mg $3s$, $3p$
complex and the other conduction bands.
Density-functional calculations within the local density
approximation and using the same Gaussian basis sets as for the
HF calculations give a gap of 5.0 eV between the valence and
conduction bands, nearly 3 eV lower than observed in experiment.

\begin{figure}[!b]
\includegraphics*[angle=90,width=0.80\columnwidth]{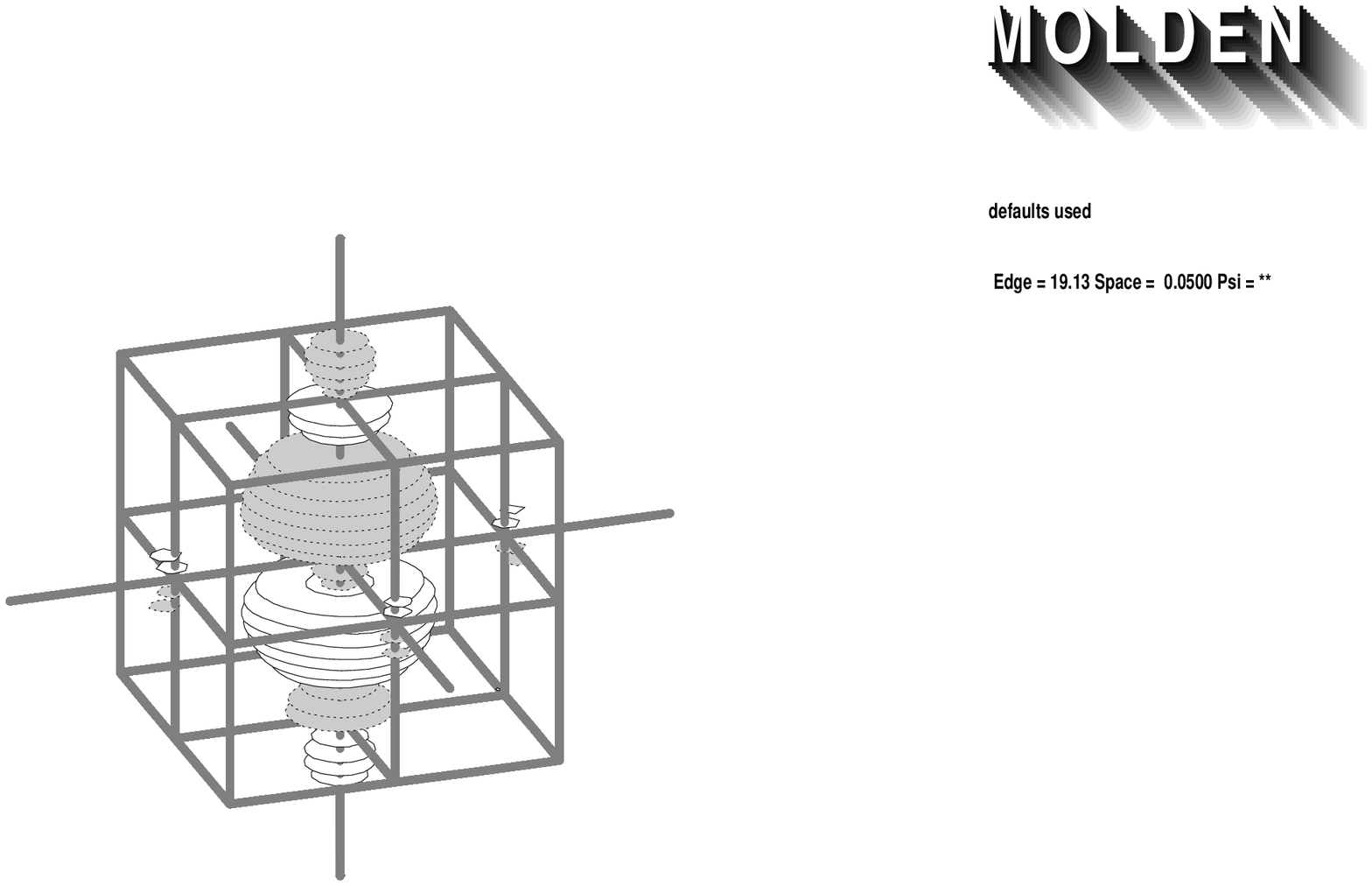}
\caption{Mg $p$-like conduction-band Wannier orbital after
projection onto a [Mg$_{19}$O$_{14}$] cluster.}
\end{figure}

Projected Wannier orbitals associated with the Mg $3s$, $3p$
conduction-band complex are plotted in Fig.2 and Fig.3.
We note that the Wannier-Boys localization module 
of the \textsc{crystal} program yields a set of Mg $sp^3$
hybrids for the lowest four conduction-band states.
In order to arrive at a set of $s$ and $x$, $y$, $z$ -like
functions, we applied to those projected hybrids a
Pipek-Mezey localization procedure \cite{pipek_mezey_89}, as 
implemented in the \textsc{molpro} package \cite{molpro_2006}.
In this localization scheme, the number of atomic orbitals
contributing to a given molecular-like composite is minimized.
Whereas the Mg $3s$ and $3p$ Wannier functions have substantial
weight at the NN ligand sites, the valence-band O $2p$ (and $2s$)
Wannier orbitals are much more compact and contributions from
neighboring ions are not visible in plots like those shown
in Figs. 2 and 3.
In the calculations reported here the norm of the projected active
orbitals is typically larger than $99.0\%$ and never below $98.5\%$
of the original crystal Wannier functions.

\subsection{Correlation induced corrections to the diagonal matrix elements}

Relaxation and polarization effects in the immediate 
neighborhood of an oxygen hole were computed by 
separate restricted open-shell HF (ROHF) calculations on
[O$_{39}$Mg$_{30}$] clusters.
The multiconfiguration MCSCF module of the \textsc{molpro} package
was employed for this purpose.
The active region $\mathcal{C}_A$ of the [O$_{39}$Mg$_{30}$] cluster
contains a central $2p^5$ (or $2s^1$) O$^-$ ion and four of the
twelve nearest O$^{2-}$ ligands.
These four oxygen neighbors are all chosen to be in the same plane
and we denote them as O$_{xy}^1$,...O$_{xy}^4$ (see Fig.4).
In addition, we include in the cluster $\mathcal{C}$ the NN cations
and all the O's in the next coordination shell of each of the
O$_{xy}^i$ sites, as sketched in Fig.4.
These additional Mg and O neighbors represent the so-called
buffer region $\mathcal{C}_B$ and ensure an accurate
description of the tails of the orbitals centered in the active
region $\mathcal{C}_A$.

\begin{figure}[b]
\includegraphics*[angle=-2,width=0.98\columnwidth]{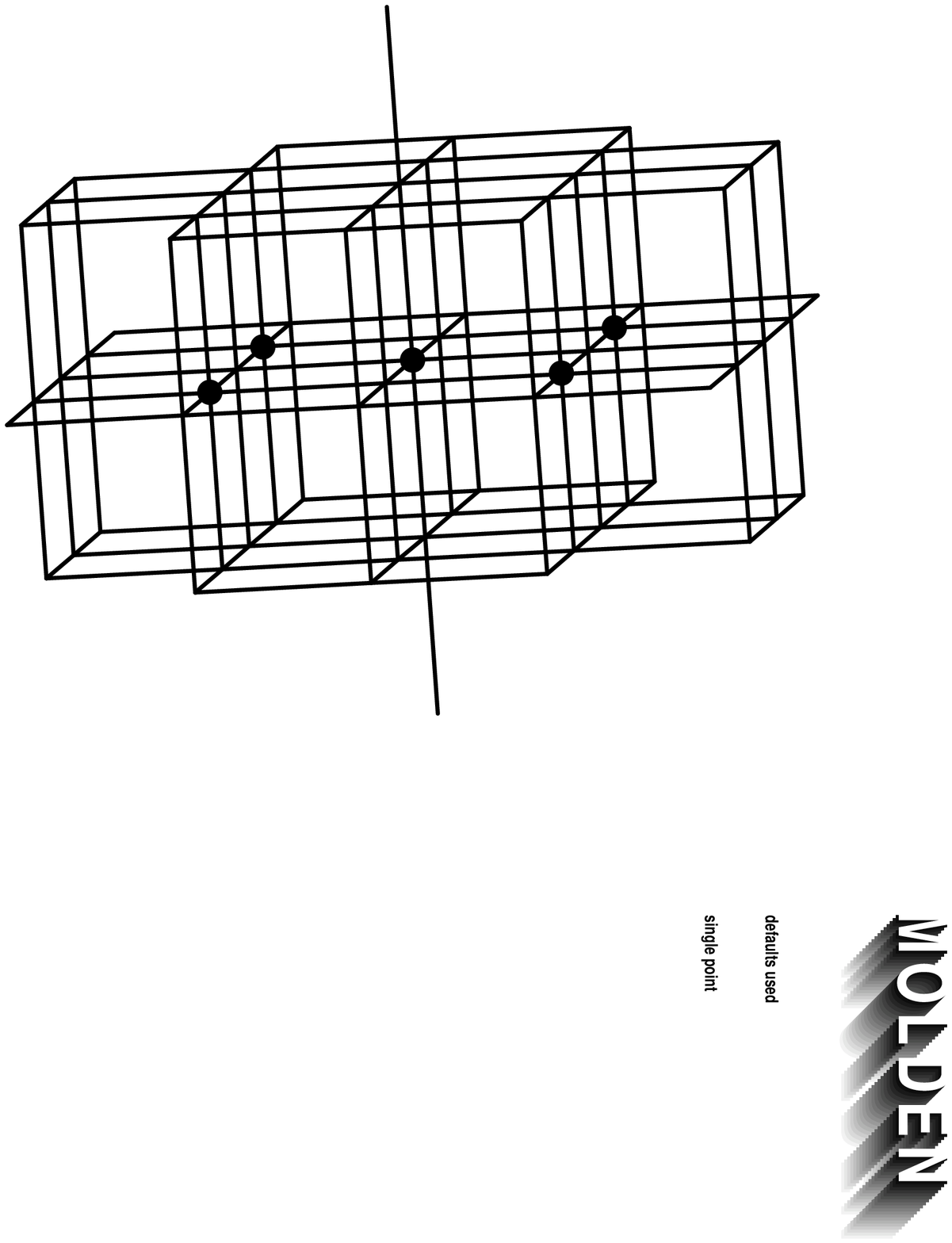}
\caption{Sketch of the [O$_{39}$Mg$_{30}$] cluster employed for
the calculation of short-range relaxation effects on
the O valence-band states.
The shortest line segments are Mg--O ``bonds''.
The so-called active region $\mathcal{C}_A$ includes a central O ion
and four nearest oxygen neighbors O$_{xy}^i$ in the ``horizontal''
plane, see text.
These active ions are shown as small black spheres.}
\end{figure}

When performing the ROHF calculations, the oxygen hole 
orbital is kept frozen \cite{CFLH_06}. 
We also freeze in our calculations the core-like $1s$,
$2s$, and $2p$ shells of all Mg ions.
Orbital relaxation effects are listed in Table I for both
$2s$ and $2p$ oxygen holes. 
The on-site relaxation effect is quite large, more than 2
eV.
There is also a substantial relaxation/polarization effect
associated with the first oxygen neighbors.
For the four O$_{xy}^i$ ligands included in the active
region $\mathcal{C}_A$ of the cluster, this effect amounts to about 0.41 eV
in the presence of an O $2s$ hole and to 0.40 eV in the
presence of a $2p$ hole.
In Table I, we multiplied these numbers by three because there
is a total of twelve oxygens in that coordination shell.
We obtain thus a good estimate for the relaxation and 
polarization effects up to the nearest O neighbors. 
The fact that these relaxation/polarization effects are
additive was checked by extra calculations with smaller,
double-zeta basis sets \cite{BSs_MgO_92} on two different 
clusters: a [O$_{39}$Mg$_{30}$] cluster including only the
four O$_{xy}^i$ sites in the active region and a 
[O$_{55}$Mg$_{38}$] cluster where all twelve first oxygen 
neighbors were allowed to polarize. 
 
The four lowest-energy $(N\!+\!1)$ conduction-band states 
imply Mg $3s^1$ and $3p^1$ electron configurations.
ROHF calculations were performed for such configurations on
a [Mg$_{19}$O$_{38}$] cluster with a [MgO$_{6}$] kernel as
active region, $\mathcal{C}_A$.
Beyond the [MgO$_{6}$] kernel, this cluster incorporates again
all Mg and O ions in the first two coordination shells of
the active ligands.
The on-site relaxation effects associated with the addition of an 
electron in a localized Mg $3s$ or $3p$ Wannier orbital are
vanishingly small.
The relaxation effects at the adjacent O sites induce energy
shifts of 0.80--0.85 eV, see Table I.
In these calculations the open-shell active orbitals
(Mg $3s^1$ or $3p^1$) were again kept frozen \cite{CFLH_06}.
We note that the energetic effect is nearly the same for the
$3s^1$ and $3p^1$ conduction-band states.
At the scale of Fig.1 at least, it induces an uniform
downwards shift of the center of gravity of the $3s\!-\!3p$ band
complex.

\begin{table}[t]
\caption{Correlation induced corrections to the diagonal
Hamiltonian matrix elements for the valence-band O $2s$, $2p$
and conduction-band Mg $3s$, $3p$ states.
All numbers are in eV.
Negative corrections induce upwards shifts of the valence bands
and shifts to lower energies for the conduction bands.}
\begin{ruledtabular}
\begin{tabular}{lrrrr}
$\Delta H_{nn}(\mathbf{0})$
                           &O $2s$    &O $2p$    &Mg $3s$    &Mg $3p$    \\
\colrule
\\
On-site orb. relaxation    &$-2.64$   &$-2.04$   &---        &---     \\
NN orb. relaxation         &$-1.23$   &$-1.20$   &$-0.81$    &$-0.84$ \\
Long-range polarization    &$-1.80$   &$-1.80$   &$-2.25$    &$-2.25$ \\
Total                      &$-5.67$   &$-5.04$
                                                 &$-3.06$
                                                             &$-3.09$ \\
\end{tabular}
\end{ruledtabular}
\end{table}

The data listed in Table I indicate that the on-site orbital
relaxation and relaxation and polarization effects at the
nearest oxygen sites in the presence of an extra electron or
extra electron hole results in a reduction of the HF band gap
by about 4.05 eV, that is, more than $45\%$ of the difference
between the HF and experimental values.
Large corrections are also expected to arise from long-range
polarization effects.
The long-range polarization energy of a dielectric due to
the presence of an extra charge $\pm e$ can be expressed
as $\Delta E(\infty ) = \Delta E(R) - C/R$ \cite{fulde_95},
where
\begin{equation}
C = \frac{\epsilon _0 - 1}{2 \epsilon _0} e^2 \ ,
\end{equation}
$\epsilon _0$ is the static dielectric constant of the material,
and $R$ defines a sphere around the extra charge beyond
which the dielectric response reaches its asymptotic value
$\epsilon _0$.
The energy increment $\Delta E(R)$ denotes the relaxation 
and polarization energy up to the radius $R$ around the
added particle.
The constant $C$ can be obtained by choosing two different
radii $R_{1}$ and $R_{2}$ where the quantities $\Delta E(R_{1})$ and
$\Delta E(R_{2})$ are computed, see for example Ref.
\onlinecite{grafen_97}.
However, we adopt here a simpler approach.
We calculate the corrections due to long-range polarization
by using the experimental value for the static dielectric
constant, $\epsilon _0=9.7$.
Since relaxation and polarization effects related to the
nearest oxygen neighbors were already accounted for, both, for the valence-band
hole states and the conduction-band electrons (see Table I) and since
the core-like electrons of the Mg$^{2+}$ ions can be ignored
in these calculations, we set $R$ as the average of the
radii of the first and second oxygen coordination shells
around a localized $2p$ hole or $3s$($3p$) electron:
$R=(a\sqrt{2}/2+a)/2$ for the O $2p$ valence-band states, 
where $a\!=\!4.217$\,\AA \ is the lattice constant \cite{mgo_LatCst_79}, 
and $R=(a/2+a\sqrt{3}/2)/2$ for the conduction-band states.
The corrections to the diagonal matrix elements of the
Hamiltonian are then
\begin{equation}
\Delta H_{nn}(\mathbf{0}) =
-\frac{\epsilon _0 - 1}{2 \epsilon _0} \frac{e^2}{R} \ ,
\end{equation}
about $-1.80$ eV for the O $2s/2p$ bands and $-2.25$ eV for the
Mg $3s$/$3p$ bands.
These numbers are also included in Table I.

Before we discuss band narrowing (or broadening) due to 
correlation effects which also affect the band gap, we consider
the loss of ground-state correlations.
The latter leads again to a shift of the center of gravity
of the bands. 
As pointed out in the previous section, some of the configurations
that are present in the $N$-particle ground-state are blocked when
an electron is added or removed.
We investigated such correlations by CI calculations with single
and double excitations (CISD) and discuss first differential
effects for the $N$ and $(N\!-\!1)$ states. 
Since the oxygen valence-band Wannier orbitals are rather
localized, we designed a cluster with a single O ion in
the active region. Around this central O site we added 
one shell of Mg ions (6 Mg's) and two shells of anions
($12\!+\!6$ O's) to build the buffer region $\mathcal{C}_B$.
In the CISD calculations for the $N$ and $(N\!-\!1)$ configurations 
we correlate the $2s$ and $2p$ orbitals of the central O ion.
Thereby the occupancy of the hole orbital is kept frozen
in the calculations for the $(N\!-\!1)$ states, which is referred
to as the frozen hole approximation \cite{birken_06,CFLH_06}.
Sets of separately optimized orbitals were used for the 
hole states of the $(N\!-\!1)$-particle system, as discussed above.
We found that for a $2p$ hole the correction to the on-site 
matrix element of the Hamiltonian is $\Delta H_{nn}(\mathbf{0})=0.85$
eV, i.e., the $2p$ valence bands are downshifted by 0.85
eV. For the O $2s$ hole states, this correction amounts to 0.99 eV.
One would expect similarly an upwards shift of the conduction 
bands.
However, the situation is somewhat different here.
When an extra electron is attached to the Mg$^{2+}$ ion, it
polarizes the closed shells of the core.
This is the dominant effect now because Mg$^{2+}$ has no valence 
electrons.
We may employ for our analysis the high-quality results obtained
for a free Mg ion by Doll \textit{et al.} \cite{doll_mgo_95}.
The correction to the ROHF Mg$^+$$\rightarrow$\,Mg$^{2+}$
ionization potential was found to be 0.27 eV in Ref. 
\onlinecite{doll_mgo_95}.
A similar differential correlation effect is occuring for
the conduction-band states in bulk MgO.
Therefore the conduction bands are shifted downwards instead of
upwards and a partial cancellation between \textit{loss of
ground-state correlation effects} for the valence and conduction
bands is taking place.
The net result is a slight increase of the gap between the
valence and conduction bands, in the range of 0.5 eV. 
Among the different contributions discussed here, this 
appears to produce the smallest corrections to the gap.
More advanced calculations for studying such
differential correlation effects are left for future work.

To summarize the results listed in Table I, relaxation and 
polarization effects in bulk MgO are responsible for a reduction
of the Hartree-Fock gap by 8.1 eV, which represents about $95\%$
of the difference between the HF and experimental values,
16.2 and 7.8, respectively.
Improved agreement is expected between our results and the
experimental data when applying higher-quality basis sets.
We mention in this context that a reduction of 3.8 eV is
obtained for the HF gap when going from valence double-zeta to
triple-zeta basis sets.
Nevertheless, this large energy difference is mainly related to the very poor
representation of the conduction-band states in the calculations
with double-zeta basis functions and such effects will be less
substantial by further extension of the basis sets.
We also keep in mind that differential correlation effects
due to the existence of a different number of electrons in
the system's ground-state and in the $(N\!\pm\!1)$ excited states
determine a small correction in the opposite direction,
i.e., a slight increase of the fundamental gap.

\begin{table}[t]
\caption{Nearest--neighbor (NN), $\mathbf{R}_{NN}\!=\!(1,1,0)a/2$,
and next-nearest-neighbor (NNN), $\mathbf{R}_{NNN}\!=\!(1,0,0)a$,
hopping matrix elements for the conduction-band Mg $3s$ and $3p$
orbitals, see text.
Results of frozen-orbital CI (FO-CI) are listed in the second
column; NOCI results in terms of separately optimized, relaxed
orbitals (RO-NOCI) are given in the third column.
All numbers are in eV.}
\begin{ruledtabular}
\begin{tabular}{lcc}
$t_{nn^\prime}(\mathbf{R})$
                        &FO-CI        &RO-NOCI \\
\colrule
$t_{NN}$\,:  \\
$3s - 3s$               &$0.41$       &$0.42$ \\
$3p_{x(y)}-3p_{x(y)}$   &$0.66$       &$0.69$ \\
$3p_{x(y)}-3p_{y(x)}$   &$0.72$       &$0.77$ \\
$3p_z - 3p_z$           &$0.13$       &$0.13$ \\

$t_{NNN}$\,:  \\
$3s - 3s$               &$0.36$       &$0.37$ \\
$3p_{x}-3p_{x}$         &$0.77$       &$0.74$ \\
$3p_{y(z)}-3p_{y(z)}$   &$0.13$       &$0.12$ \\
\end{tabular}
\end{ruledtabular}
\end{table}

\subsection{Off-diagonal matrix elements}

We discuss next the effect of correlations on the widths of
the different bands.
For that purpose the off-diagonal matrix elements of the effective
Hamiltonian (\ref{eqHkcorr}) 
have to be determined, i.e., the so-called hopping terms.
At the Hartree-Fock level, these matrix elements are obtained by
solving $2\!\times\!2$ secular equations where both wavefunctions are 
expressed in terms of localized HF orbitals, see Eqs.~(\ref{eqPhiNpm1}), 
(\ref{eqHkHF}), and (\ref{eqtnmHF}). 
Relaxation and polarization effects in the nearby 
surroundings of the added electron (or hole) are obtained by separate 
SCF optimizations for the $(N\!\pm\!1)$ states.
The separate optimization of the $(N\!\pm\!1)$ wavefunctions leads to
sets of non-orthogonal orbitals.
There will be thus both Hamiltonian and overlap matrix elements
between the extra electron (extra hole) wavefunctions
$\Psi_{\mathbf{0}n\sigma}^{N\pm 1}$ and
$\Psi_{\mathbf{R}n^\prime\sigma}^{N\pm 1}$.
The calculation of such matrix elements has been recently implemented
in \textsc{molpro} \cite{SI_molpro}.
It is based on the transformation of the CI vectors to bi-orthogonal
orbitals and follows an idea suggested by Malmqvist \cite{SI_Malmqvist_86}.
A similar approach was developed by Broer \textit{et al.} in
Groningen \cite{NOCI_Broer_81_88}.
Instead of putting these matrix elements directly into the eigenvalue
equations that determine the Bloch energies $\epsilon_n(\mathbf{k})$
we extract from these data effective hopping parameters associated
with various pairs of orbitals.
Comparison between such effective hopping terms and the HF off-diagonal
Hamiltonian matrix elements offers an insightful picture of how 
correlation effects modify the inter-site interactions and
consequently the widths of the bands.
For energetically degenerate states, the effective hopping 
is defined as  
\begin{equation}
t_{nn^\prime} = (H_{nn^\prime}
   - S_{nn^\prime}H_{nn})/(1-S_{nn^\prime}^2) \,,
\end{equation}
where $H_{nn^\prime}$ and $S_{nn^\prime}$ are the Hamiltonian
and overlap matrix elements between the $(N\pm 1)$ 
states $n$ and $n^\prime$.
Since the separately optimized wavefunctions are expressed in
terms of sets of non-orthogonal orbitals, this type of secular
problem is usually referred to as non-orthogonal CI (NOCI).
In the case of mutually orthogonal states $S_{nn^\prime}$ is zero 
and $t_{nn^\prime}(\mathbf{R}) = H_{nn^\prime}(\mathbf{R})$.

\begin{figure}[t]
\includegraphics*[angle=90,width=0.85\columnwidth]{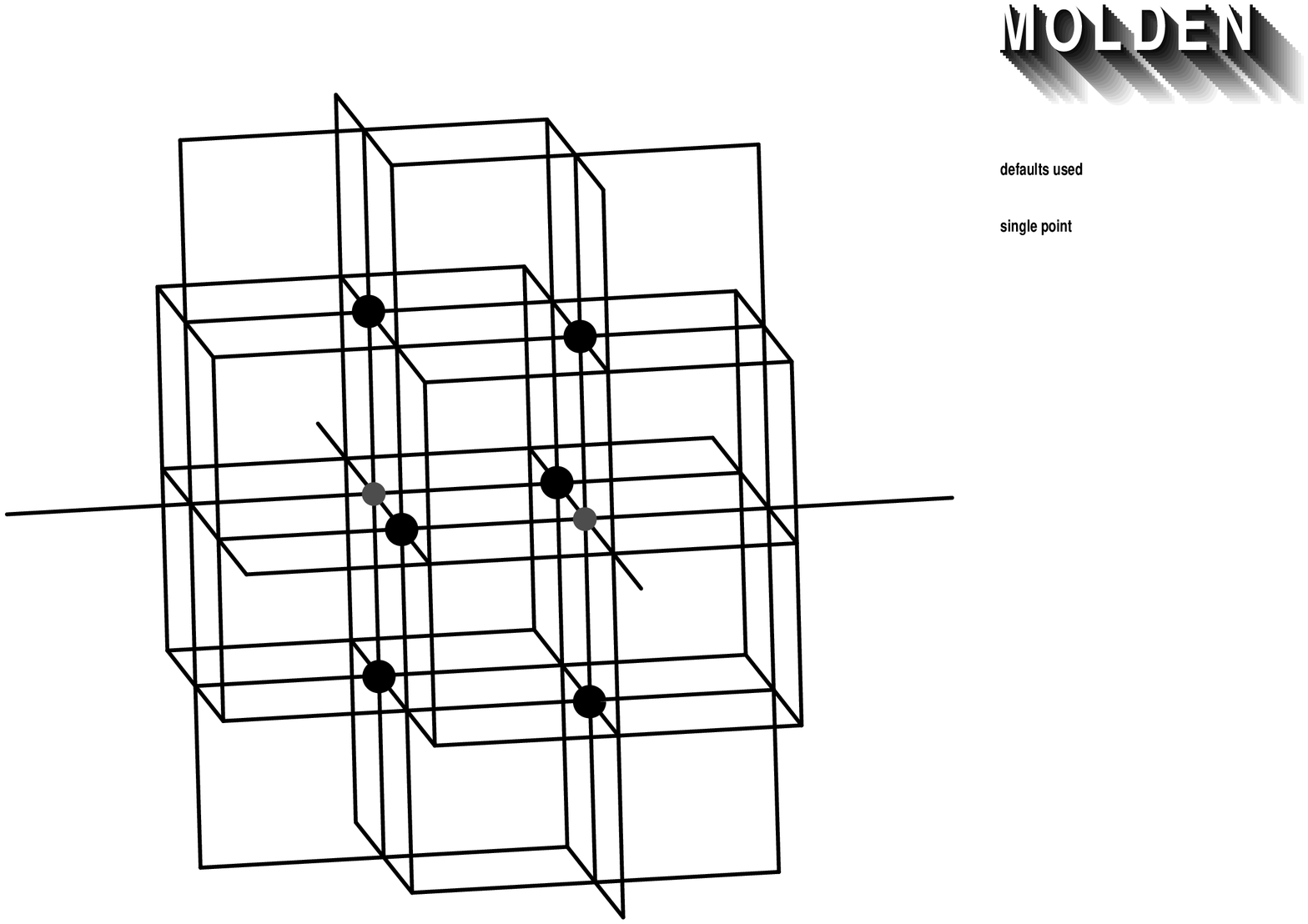}
\caption{Sketch of the [Mg$_{28}$O$_{36}$] cluster employed for
the calculation of the NN conduction-band hopppings.
The active region $\mathcal{C}_A$ includes two NN Mg sites, small
grey spheres, plus the bridging and apical ligands, large black
spheres.}
\end{figure}

Nearest-neighbor and next-nearest-neighbor (NNN) hopping matrix
elements for the more diffuse conduction-band Mg $3s$ and $3p$ 
orbitals are listed in Table II.
In the rocksalt structure, there are two O ions bridging
two NN cations, whereas two NNN Mg's share a single oxygen.
We designed a [Mg$_{28}$O$_{36}$] cluster for calculating
the NN hoppings and a [Mg$_{36}$O$_{47}$] cluster for the NNN
matrix elements.
As active regions, we employed [Mg$_{2}$O$_{6}$] and
[Mg$_{2}$O$_{9}$] kernels, respectively, see Fig.5 and Fig.6.
As in the calculations for the on-site matrix elements, for
each of these clusters we included in the buffer region
$\mathcal{C}_B$ all metal and O ions in the first two coordination
shells of the active oxygens.
Since the second cluster consists of 83 atoms, we reduced
the computational effort by removing the polarization
functions at the oxygen sites for this particular cluster.
Two different values are given in Table II for each hopping
matrix element.
Results extracted from $2\!\times\!2$ CI calculations in terms of
frozen HF Wannier orbitals are listed in the second column. 
In the third column, we allowed for full relaxation of the $2s$
and $2p$ orbitals of the O ions included in the active region
$\mathcal{C}_A$, for each of the $(N\!+\!1)$ configuration state 
functions entering the $2\!\times\!2$ CI.
As illustrated in Figs. 5 and 6, those active anions are the ligand(s) bridging two Mg
sites and the ligands which are nearest oxygen neighbors of one
of the active Mg sites and also of the bridging ligand(s).

The results show that the separate
optimization of the $(N\!+\!1)$ wavefunctions induces only
minor changes on the electron hoppings, in the range of
few percent.
An interesting feature is that variations occur in both
directions, i.e., some of the effective hoppings are slightly
enlarged by taking into account short-range relaxation and
polarization effects and some are reduced.
Since these changes are quite small, the width (and the overall
structure) of the lower conduction bands will change very little.
Regarding the longer-range polarization effects, we expect that
their influence on the hopping terms is negligible.

In the case of non-equivalent orbitals, neighboring Mg $3s$
and Mg $3p$, it is more difficult to define some NOCI effective
hoppings and the corresponding data is not shown in Table II. 
Nevertheless, the effect of short-range relaxation and
polarization is also small for these interactions, with reductions
of the CI splittings of few meV.

\begin{figure}[b]
\includegraphics*[angle=90,width=0.85\columnwidth]{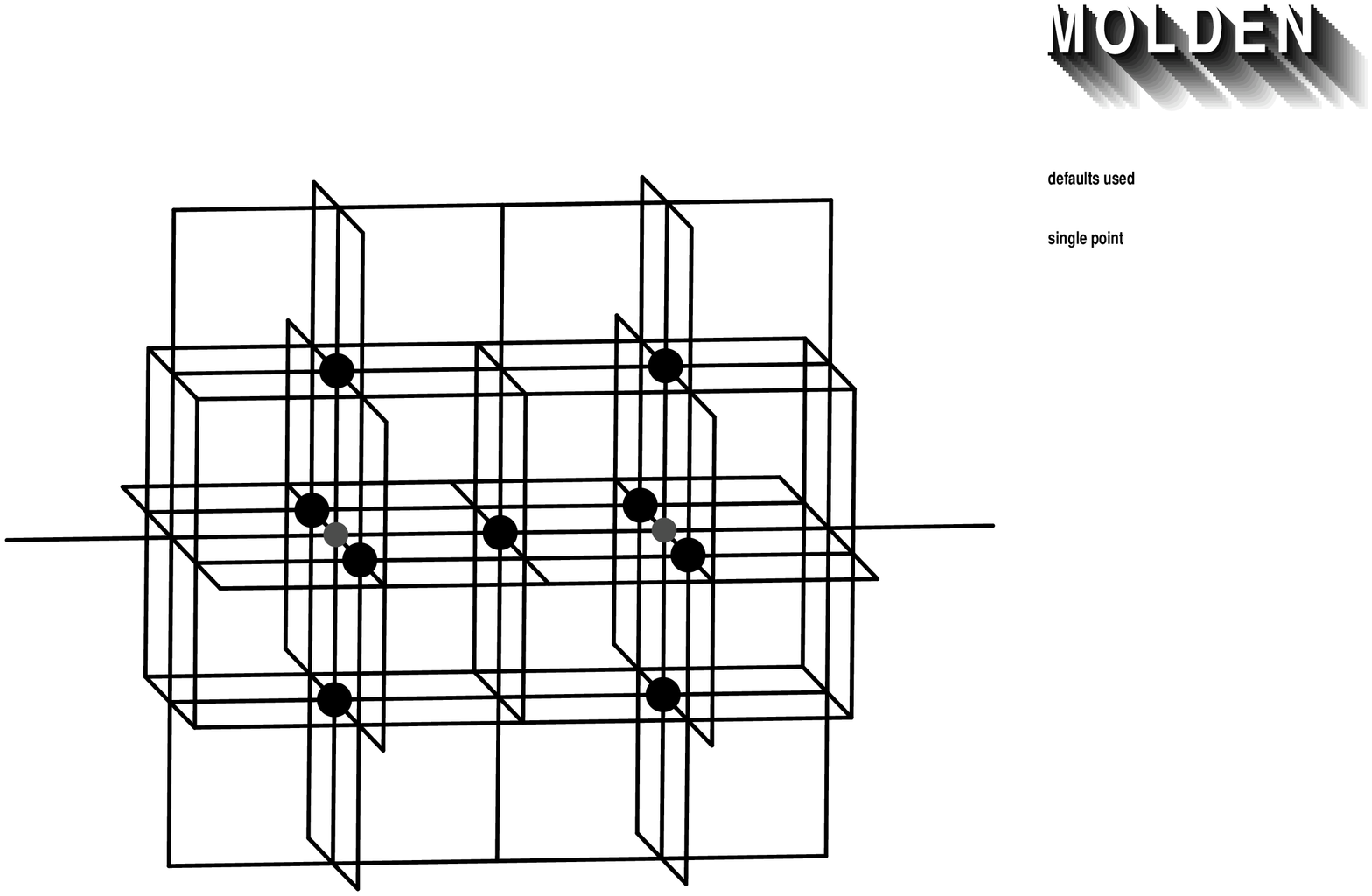}
\caption{Sketch of the [Mg$_{36}$O$_{47}$] cluster employed for
the calculation of the NNN conduction-band hopppings.
The active region $\mathcal{C}_A$ includes two Mg sites, small
grey spheres, and nine O neighbors, large black spheres.}
\end{figure}

The same type of analysis was applied to the $(N\!-\!1)$ O $2s^1$/$2p^5$ 
hole states.
The spatial extent of the oxygen orbitals and the inter-site matrix
elements are significantly smaller.
Since the largest relaxation effects concern orbitals at the same
site, see Table I, we included in a first step only two ligands in the
active regions of our clusters.
Nevertheless, few other atomic shells were added around these active
sites in the buffer region $\mathcal{C}_B$.
We considered a [O$_{28}$Mg$_{10}$] cluster for computing the 
relaxation effects on the NN hoppings and a [O$_{32}$Mg$_{11}$] cluster
for the NNN terms.
The results are collected in Table III.
The largest corrections arise for the NN matrix elements, with 
absolute values that are similar to the corrections obtained for the
NN conduction-band hoppings. 
In relative numbers, these corrections are somewhat larger, by 
$10\%$ to $20\%$ for the NN hoppings.
A very different situation occurs in strongly correlated
oxides such as the layered cuprates, where the existence of an
antiferromagnetic spin background determines a reduction of the
effective quasiparticle hoppings by a factor of four \cite{ZR_hopp_06}.

The trends displayed in Table III are confirmed by multi-reference 
CISD [MRCI(SD)] calculations.
We correlated in these calculations the eight $2s$ and $2p$ orbitals
at the two NN oxygen sites and the reference active space included 
those two orbitals involved in the hopping process.
The hopping matrix element is half of the energy separation between
the lowest two eigenstates. 
The MRCI values for the $2p_x\!-\!2p_x$ and $2p_x\!-\!2p_y$ hoppings,
for example, are 0.33 and 0.43 eV.
These numbers are again larger than the HF results, although the
magnitude of the effect is smaller as compared to the NOCI calculations.
Having the data of the SCF calculations for the $(N\!-\!1)$ oxygen hole states at hand,
we performed an analysis of the changes produced in the composition
of the (relaxed) orbitals in the immediate vicinity of an O$^-$ $2s^1$
or $2p^5$ anion. 
We found that an oxygen hole is causing polarization and 
``bending'' of the $2p$ orbitals at the NN ligand sites.  
The ``bending'' of the NN $2p$ orbitals towards the O$^-$ site takes
place through both $p\!-\!s$ and $p_i\!-\!p_j$ mixing.
These effects result in stronger inter-site orbital overlap and
explain the fact that the effective NOCI hoppings are \textit{larger} than
the corresponding HF values. 
A similar analysis for the conduction-band $(N\!+\!1)$ states is complicated by
the presence of several sets of ``active'' orbitals, Mg $3s$, $3p$ and bridging
O $2p$, and we could draw no clear conclusions in that case.

\begin{table}[!t]
\caption{NN, $\mathbf{R}_{NN}\!=\!(1,1,0)a/2$,
and NNN, $\mathbf{R}_{NNN}\!=\!(1,0,0)a$,
hopping matrix elements for the valence-band states.
Results of frozen-orbital CI (FO-CI) are listed in the second
column; NOCI results in terms of separately optimized, relaxed
orbitals (RO-NOCI) are given in the third column.
All numbers are in eV.
Data for the oxygen $2s$ orbitals are also included, although
the O $2s$ band is much below the O $2p$ bands.
The values in parentheses include orbital relaxation effects at four
additional O sites, see text. 
}
\begin{ruledtabular}
\begin{tabular}{lcl}
$t_{nn^\prime}(\mathbf{R})$
                        &FO-CI     &RO-NOCI    \\
\colrule
$t_{NN}$\,:  \\
$2s - 2s$               &$0.10$    &$0.12$  \ ($0.11$)   \\
$2p_{x(y)}-2p_{x(y)}$   &$0.32$    &$0.37$  \ ($0.36$)   \\
$2p_{x(y)}-2p_{y(x)}$   &$0.42$    &$0.49$  \ ($0.47$)   \\
$2p_z - 2p_z$           &$0.12$    &$0.14$  \ ($0.13$)   \\

$t_{NNN}$\,:  \\
$2s - 2s$               &$0.01$    &$0.01$   \\
$2p_{x}-2p_{x}$         &$0.06$    &$0.06$   \\
\end{tabular}
\end{ruledtabular}
\end{table}

Extra calculations were performed for the NN hopping matrix elements
with four additional oxygen sites included in the active region of the
cluster, $\mathcal{C}_A$.
Those are the four ligands which are nearest neighbors to both O ions
involved in the hopping process.
They are situated in the median plane of the segment $\mathbf{R}_{NN}\!=\!(1,1,0)a/2$.
A cluster composed of 62 sites, [O$_{40}$Mg$_{22}$], was employed for
these calculations.
The results are given in Table III in parentheses.
The corrections due to relaxation and polarization at the four
nearest O neighbors are very small, 0.01 to 0.02 eV.
The effect of these corrections is to reduce somewhat the absolute values
of the NN hoppings.

Experimental studies for characterizing the valence electronic structure
of MgO have been carried out using x-ray photoelectron spectroscopy
(XPS) \cite{mgo_kowalczyk_77} and angle-resolved ultraviolet photoelectron
spectroscopy (ARUPS) \cite{mgo_tjeng_90}.
The measured width of the O $2p$ bands is about 6.5 eV \cite{mgo_kowalczyk_77,mgo_tjeng_90}. 
The HF valence-band width is 5.50 eV and inclusion of local 
correlations leads to a slight broadening of the O $2p$ bands, 
which brings the \textit{ab initio} result in good agreement with the
experiment.
For comparison, density-functional calculations within the local density
approximation and using the same Gaussian basis sets as in the
HF calculations predict a width of 4.68 eV for the O $2p$ bands.
 
The correlation induced corrections to the widths of the bands also
influence the band gap.
In the \textsl{fcc} lattice the dispersion of $p$ bands at the $\Gamma$ point
depends on two of the nearest-neighbor hoppings,
$\epsilon_x(\Gamma) = const. + 8t_{x,x}^{(110)} - 4t_{x,x}^{(011)} + ...$
\cite{SlaterKoster_54,ashcroft_mermin}.
With the notations from Table III, $t_{x,x}^{(011)} = t_{z,z}^{(110)}$. 
Corrections of 0.04 eV for $t_{x,x}^{(110)}$ and 0.01 eV for
$t_{x,x}^{(011)}$ (or $t_{z,z}^{(110)}$), see Table III, imply an upwards shift of
the O $2p$ bands at the $\Gamma$ point and a narrowing of the fundamental 
gap by about 0.30 eV.
For the Mg $3s\!-\!3p$ conduction-band complex such changes at the
$\Gamma$ point are smaller because the correlation induced corrections
to the $3s\!-\!3s$ and $3s\!-\!3p$ inter-site matrix elements are 
lower.

\section{Summary and conclusions}

We have analyzed the different correlation contributions to the energy
gap of MgO and to the widths of the conduction and valence bands.
This was done within the quasiparticle description.
As regards correlation effects we have distinguished between relaxation
and polarization around an electron or hole added to the (neutral)
ground-state.
The net result is a reduction of the Hartree-Fock gap from 16.2 eV to
a value of 8.1 eV.
This has to be compared with a measured energy gap of 7.8 eV.
Within the local density approximation (LDA) to density functional
theory a gap of 5.0 eV is found. This is not surprising since LDA is
known to produce too small gaps for insulators.

The calculations were performed with triple-zeta basis sets.
Since on the Hartree-Fock level the calculated gap differs for double-
and triple-zeta basis sets by 3.8 eV, one may consider the good
agreement with the experimental gap as somewhat fortuitous.
Triple-zeta basis sets are known, however, to produce reliable results
in quantum chemistry and therefore a further extension of the basis
set should keep the corrections small.
Presently we are not able to work with larger basis sets.

It was shown that a large contribution to the correlation induced
corrections to the fundamental gap comes from on-site and nearest-neighbor
relaxation, i.e., from the immediate neighborhood of the added particle.
But also the long-range part of the polarization generated by the extra
particle contributes substantially to the reduction of the HF gap.
This long-range part can be treated in a continuum approximation thereby
using the known dielectric constant of MgO.
The so-called loss of ground-state correlations makes a small contribution
in MgO.
The reason is that the conduction-band Wannier orbitals have predominant
Mg $3s$ or $3p$ character.
The added electron will essentially go thus to a Mg site where it polarizes 
the closed $1s^2$, $2s^2$, and $2p^6$ shells. 
This effect reduces the gap and counteracts the loss of ground-state 
correlations which occurs when an electron is removed (hole state) and
therefore can no longer contribute to the correlations of the remaining 
ones.
Finally, also changes in the widths of the bands influence the energy
gap.

One surprising effect which we found is an enhancement of the width
of the valence bands when local correlations are taken into account.
This is the opposite one expects since correlations lead usually to
band narrowing instead of broadening.
It results from a slight bending of nearest-neighbor $2p$ orbitals
towards a O$^-$ site when we account for correlations.
The bending of the nearest-neighbor ligand orbitals increases the
wavefunction overlap and hence facilitates the hopping. 

The present results prove the usefulness of band calculations based
on quantum chemical techniques.
They allow for well controlled approximations and a transparent interpretation
of the different microscopic processes which determine the size of the
energy gap and the widths of the bands.

\

We thank Dr. E. Pahl and Prof. J. Fink for fruitful discussions.

\end{document}